\documentstyle[prl,aps,twocolumn,psfig]{revtex}
\begin{document}
\draft
\title{THEORETICAL APPROACH TO THE CORRELATIONS OF A HARD-SPHERE CRYSTAL}

\author{C. Rasc\'{o}n$^{1,2}$, L. Mederos$^2$, and G. Navascu\'{e}s$^1$}
\address{$^1$Departamento de F\'{\i}sica Te\'{o}rica de la Materia
Condensada, Universidad Aut\'{o}noma, Cantoblanco, Madrid E-28049,Spain}
\address{$^2$Instituto de Ciencia de Materiales (Consejo Superior de
Investigaciones Cient\'{\i}ficas), Cantoblanco, Madrid E-28049, Spain}
\date{\today}
\maketitle

\begin{abstract}
We present the first theoretical approach to the angular-average of the
two-body correlation function $\tilde g(r)$ for simple solids. It is based
on three sum rules for $\tilde g(r)$: the compressibility and virial
equations and the normalization. We apply the theory to determine this
correlation function for the case of the FCC solid phase of hard spheres. The
agreement with simulation data is excellent over all the density range. The
application to other simple systems is discussed. The approach opens a new
route to perturbation theories for simple solids.
\end{abstract}


The most important correlation function in fluid classical systems is the
radial distribution function $g(r)$. The theoretical accessibility of this
function for a hard-sphere (HS) fluid, namely the Percus-Yevick
approximation, has been the crucial point to the extraordinary development of
the theory of simple liquids \cite{PY,HM}. The analogous progress is lacking
in the case of
 structured phases of these systems, the solids. Spite of the advance
in the theory of classical non-uniform systems experimented during the last
fifteen years \cite{Hen}, one of the most important theoretical objectives,
correlations, has remained unapproachable.

In structured systems the most important correlation is $\tilde g(r)$ the
angular average of the two-particle density function $\rho^{(2)}({\bf r},{\bf
r}')$. It appears in all crucial equilibrium equations as the virial and
compressibility equations or the energy equation \cite{HM} and it is the key
of all perturbation approaches \cite{HM,Hen,Weeks}. The $\tilde g(r)$ plays
the same rule in classical solids as $g(r)$ does in classical fluids.
However, up to now, no theoretical approach to $\tilde g(r)$ has been
reported. Here, we present the first theoretical approach to this function
and apply it to evaluate $\tilde g(r)$ for a face-centred-cubic HS solid. The
agreement with simulation results is excellent over all the physical range of
densities from below melting up to almost close packing.

All previous theoretical approaches reported in the literature to the
correlations of classical solids have been focused on $g({\bf r},{\bf r}')$
(defined through $\rho^{(2)}({\bf r},{\bf r}')\equiv\rho({\bf r})\rho({\bf
r}')g({\bf r},{\bf r}')$) for HS and mainly in relation to perturbation
schemes. These are not true approximations, rather, they are simple mappings
to the radial distribution function of a uniform system at some effective
density. None of these approaches can resist a direct comparison with
simulation results \cite{Kirlidis}. In fact, these mappings could be done
over a wide family of functions \cite{M94}. Moreover, only $\rho^{(2)}({\bf
r},{\bf r}')$ (the probability to find two particles at ${\bf r}$ and ${\bf
r}'$) and $\rho^{(2)}({\bf r},{\bf r}')/\rho({\bf r})$ (the probability to
find a particle at ${\bf r}$ provided there is another one at ${\bf r}'$) and
their angular averages have a direct physical meaning. Then it would be more
sensible to make approximations on these functions than on  $g({\bf r},{\bf
r}')$. The function $\tilde g(r)$ is defined as
\begin{equation}
\tilde g(r) =  {{1} \over {4 \pi V \rho^2}}
\int d \Omega \int d{\bf r}_1 \rho^{(2)}({\bf r}_1,{\bf r}_1+{\bf r}),
\label{defg}
\end{equation}
where $V$ is the volume, $\rho \equiv N/V$ is the mean density
and $d \Omega$ the differential solid angle aperture around ${\bf r}$.

To motivate the new theory we shall first discuss the two types of
correlations that $\tilde g(r)$ must accounts for in a solid. The long-range
correlation characteristic of the structured phases is directly due to the
periodicity of the solid. If no other correlation is considered each particle
would move around a lattice site independently of each others. The
probability of finding a particle at ${\bf r}$ is given by the local density
$\rho ({\bf r})$ which is a sum of Gaussian-like functions located at each
lattice site. Then $\rho^{(2)}({\bf r},{\bf r}')$ is given by the simple
product of individual probabilities $\rho ({\bf r})\rho ({\bf r}')$
multiplied by a step-like function, $g({\bf r},{\bf r}')$, to avoid the
double occupancy. Thus, the long-range correlation {\it is already} described
by the product $\rho ({\bf r})\rho ({\bf r}')$. Its proper angular average
gives the long-range contribution to $\tilde g(r)$. Then, it is 
convenient to define $\tilde g_0 (r)$
 as the angular average of the product $\rho ({\bf r})\rho ({\bf r}')$,

\begin{equation}
\tilde g_0 (r) =  {{1} \over {4 \pi V \rho^2}}
\int d \Omega \int d{\bf r}_1 \rho({\bf r}_1)\rho({\bf r}_1+{\bf r}),
\label{defg0}
\end{equation}
which has the form of a sum of Gaussian-like peaks
${\tilde g_0}^{(i)}$ centred around successive neighbours distances
$R_i$, normalized to the corresponding number of neighbours $n_i$.
Then, if no other correlation than the long-range correlations imposed by the
lattice is considered, ${\tilde g}^{(i)}={\tilde g_0}^{(i)}$ for $i>0$, and
both functions are the same except that the first peak of $\tilde g_0 (r)$ is
excluded in $\tilde g(r)$.  This long-
range correlation is the only one existing in the close-packing limit of a HS
solid. In this limit $\rho^{(2)}({\bf r},{\bf r}')$ is exactly a sum of
products of $\delta$ functions at any pair of different lattice sites
multiply by a step-like function to exclude the double occupancy. Hence, the
function $\tilde g(r)$ will be a sum of $\delta$ functions at different
neighbours distances, except at zero distance, normalized to the
corresponding number of neighbours.

To take into account other correlations the successive peaks ${\tilde
g}^{(i)}$ must be increasingly modified from the ${\tilde g_0}^{(i)}$ as $i$
decreases to $1$. The compressibility equation \cite{HM}

\begin{equation}
{4 \pi \rho}
\int dr {r}^2 [\tilde g(r)-\tilde g_0 (r)]=
- 1 + \rho k_{B} T \chi_{T}.
\label{compre}
\end{equation}
is a sum rule for $\tilde g(r)$ which help us to understand these
modifications. The integral of $\tilde g_0(r)$ in Eqn.(\ref{compre}) gives
the number of particles inside the system provided the origin is located at a
lattice site. The integral of $\tilde g(r)$ gives the number of particles
inside the system (minus one because of the self-exclusion) provided a
particle is fixed at the origin. Thus the left hand side of
Eqn.(\ref{compre}) can be understood as the number $\Delta N$ of particles
(minus one) coming into the system when a particle is fixed at the origin,
therefore $\Delta N=\rho k_{B} T \chi_{T}$. Observe that in the HS close-
packing limit $\chi_{T}=0$, the correlation reduces to the long-range one
discussed above, hence ${\tilde g}^{(i)}\equiv{\tilde g_0}^{(i)}$ and
Eqn.(\ref{compre}) is verified identically. In other words, no HS come into
the system if it is completely packed. If we imagine a spherical system of
radius $R$ we can estimate $\Delta N$ as $\rho 4 \pi R^2 \delta R$, where
$\delta R$ is the displacement of the peaks at the border of the system, {\it
i.e.} the displacement of ${\tilde g}^{(i)}$ respect to ${\tilde g_0}^{(i)}$
for $R_i \approx R$.
Then $\delta R \approx \rho k_{B} T \chi_{T}/ (\rho 4 \pi R_i^2)$.
This is a quite interesting result which shows that the differences,
which should include some kind of deformation, between ${\tilde g}^{(i)}$ and
${\tilde g_0}^{(i)}$ reduce quadratically with the distance.

Until here the discussion has been quite general and now we apply it to the
case of a HS solid. The compressibility of the HS solid is so small that
$\delta R$ would be practically imperceptible in a simulation. A rough
estimation of this displacement is already negligible for the first peak even
at the lowest densities: taking $R \approx 1$ and $\rho k_{B} T \chi_{T}
\approx 0.02$ give $\delta R \approx 0.001$ (distances in HS diameter units
$d_{HS}$). Thus the location of the peaks of $\tilde g(r)$ does not differ
from the localization of those of $\tilde g_0 (r)$. If the displacement of
the peaks is negligible it is quite sensible to assume that
their functional form
cannot differ significatly from the peaks of $\tilde g_0 (r)$ except for
the first one due to the characteristic cutoff of the HS.

Up to here we have described the subjacent physics which motivates the
following theoretical approach to $\tilde g(r)$. Accordingly we approximate
the peaks of $\tilde g(r)$ by those of $\tilde g_0(r)$ beyond nearest-
neighbours. To go far toward an explicit form for these peaks, we can regard
the existent functional theories \cite{Hen} which give an accurate
description of the free energy of a HS solid and from which it is possible to
determine $\rho({\bf r})$. All these theories use successfully the Gaussian
parametrization of $\rho({\bf r})$ \cite{alfa},

\begin{equation}
\rho({\bf r}) = {({{\alpha}\over{\pi}})^{3\over2}} \sum_{\bf R_i}
{e^{-\alpha {({\bf r}-{\bf R}_i})^2}},
\label{gauss}
\end{equation}
where $\alpha$ is the Gaussian width parameter. Using Eqn.(\ref{gauss}) into
Eqn.(\ref{defg0}) yields:

\begin{equation}
{\tilde g_0}^{(i)}(r) = 
{{1} \over {4 \pi \rho}} {({{\alpha} \over {2 \pi} } )^{1 \over
2}}{n_i}{{e^{-\alpha(r-R_i)^2/2}}+{e^{-\alpha(r+R_i)^2/2}}
\over {r R_i}}, \ \ \ \ \ \ i > 0.
\label{giga}
\end {equation}
For the sake of simplicity, we have dropped all terms which arise from the
exponential products with ${\bf R}\neq{\bf R'}$ in (\ref{giga}). At the usual
values of $\alpha$ they give negligible contributions because of the absence
of overlapping.

The above 
discussion on the compressibility suggests for ${\tilde g}^{(1)}(r)$ a
functional form similar to that of Eqn.(\ref{giga}). We propose the simple
parametric form:

\begin{equation}
{\tilde g}^{(1)}(r) =
{{ A {e^{-\alpha_1(r-r_1)^2/2}}}\over {r}} \ \ \ \ \ \ r \geq d_{HS},
\label{g1}
\end {equation}
with ${\tilde g}^{(1)}(r) =0$ for $r \leq d_{HS}$. The compressibility shows
that even in the less favourable case, at the lowest densities, the
displacement of the first peak is quite small. This suggests that mean
location of the nearest-neighbours $<r>$ can be approximated by the mean value
obtained with the first peak of $\tilde g_0 (r)$:

\begin{equation}
{4\pi\rho \over n_1 }<r> \equiv {\int d{\bf r} r {\tilde g^{(1)}(r)}} =
{\int d{\bf r} r {\tilde g_0^{(1)}(r)}}. 
\label{r1media}
\end{equation}
This equation is a
sum rule for $\tilde g(r)$ becomes less approximate as the mean density
increases and it is exact at the limit of close-packing. Besides two other
{\it exact} sum rules must be obeyed by $\tilde g(r)$. The first one
corresponds to the normalization of ${\tilde g}^{(1)}$ to the
nearest-neighbours number:
\begin{equation}
4 \pi \rho \int_{d_{HS}}^{\infty} dr r^2 {\tilde g}^{(1)}(r) = n_{1}.
\label{norma}
\end{equation}
The virial equation is the second {\it exact} sum rule. It can be easily
proved that, for non-uniform systems, the pressure is related to the value
of $\tilde g(r)$ at contact exactly in the same way as it is related to the
radial distribution function of uniform fluids:

\begin{equation}
\beta P / \rho = 1 + 4 \eta {\tilde g}(d_{HS}),
\label{presion}
\end{equation}
where $\beta = 1/k_BT $ and $\eta$ is the packing fraction ($\eta
=\pi\rho/6$).

All the required information to determine $\tilde g(r)$ ($\alpha$ and
pressure as functions of $\rho$) is now provided by the minimization
of any of the well known and accurate density functionals for the Helmholtz
free energy of HS solid. With these data, the three proposed sum rules form a
non-linear system of equations which is solved to find $A$, $\alpha_1$ and
$r_1$ at each $\rho$. Simultaneously, using $\alpha$ and
Eqn.(\ref{giga}), the successive peaks of ${\tilde g(r)}$ are obtained. Very
recent studies have shown that the equation of state of the HS solid deduced
from different functional approaches agree quite well with simulation results
over all the density range \cite{eoe}. For the following calculation, we use
the generalized effective liquid approximation (GELA) \cite{GELA} as it gives
the best overall behaviour. However, there is not significant differences if
any other functional approach is used. If we compare the most recent Monte
Carlo simulations by Choi {\it et al.} \cite{Choi91} with the predictions of
the present theory for $\tilde g(r)$, the agreement is excellent over all
densities and specially impressive at high densities. Figures 1 and 2 show
$\tilde g(r)$ for two significant densities: $\eta=0.52$, the lowest
density below melting ($\eta_m\approx0.54$) with available simulation data,
and $\eta=0.71$, near close packing ($\eta_{cp}\approx0.74$), respectively.
There are some differences between theoretical predictions and simulation
results. Nevertheless, they are quite small and can only be appreciated more
easily at the lowest densities. Let us first pay attention on ${\tilde
g}^{(1)}(r)$. The value at contact differs from that of simulation. It is a
direct consequence, via virial equation, of the approximate theoretical
pressure. If the {\it exact} pressure (from simulation) is used into the
theory, the agreement with simulation would be almost complete (see figure 1)
confirming the goodness of the theory. The rest of the peaks of $\tilde
g(r)$, which only depends on the parameter $\alpha$ ($\alpha=113$ for
$\eta =0.54$ and $\alpha =10094$ for $\eta =0.71$), also agree quite well with
the simulation results. Introducing again the {\it exact} $\alpha$ values
(estimated from simulation: $\alpha=91$ for $\eta=0.54$ and
$\alpha=7659$ for $\eta=0.71$) the agreement is excellent. The Gaussian
parameter $\alpha_1$ is approximately half than those of the rest of the
peaks ($\alpha_1=50$ for $\eta=0.54$  and $\alpha_1=5405$ for $\eta=0.71$).
However,because the cutoff the width of this first peak is similar to the
rest. The excellent agreement of the first peak showed in the inset of the
figures would give an estimation of $\alpha_1$ from simulation practically
equal to that predicted by the theory. More interesting is the $r_1$
parameter, it corresponds to the position where the
first peak has its practically maximum value.
 Notice, however, that at very low densities
the real maximun is located at contact (see figure 1) and $r_1$ would be
the maximun if the first peak is analytically extrapolated below $r=1$.
The important point is that simulation data of this maximum has been
reported (at very low densities the extrapolation has been also reported).
Figure 3 shows the these data for different densities together with
the theoretical predictions.

  According with the decreasing of the
lattice parameter, the width of the first peak decreases monotonously with
$\rho$ ($\alpha_1$ increases monotonously with $\rho$).

$r_1$ must be always smaller that the $<r>$ except at the close-packing
limit where both coincide. As $\rho$ decreases from this limit the
pressure decreases rapidily from infinity and therefore the value at contact.
Meanwhile the lattice parameter hardly changes
 and the peak width still remains quite sharp. under
these circumstances the only way to keep on with the normalization is
increasing $r_1$. However, at low $\rho$, the pressure does not change
too much with $\rho$ and the peak becomes duller as $\rho$
decreases, $r_1$ must recede to maintain the normalization. We remark this
because the overall agreement of the peaks does not necessarily implies the
nice agreement of the behaviour of $r_1$ with the mean
density. This proves the suitable physical description of our theoretical
approach. The parameter $A$ is a simple factor to adjust the normalization or
to adjust the value at contact.

A major consequence of the theoretical knowledge of $\tilde g(r)$ is the
possibility to develop and use proper perturbation theories for solids where
the perturbative term uses the appropriate correlation function of the
reference HS system instead of a the correlation of a fluid at same effective
density \cite{Hen}.  Working along this direction is in progress.
Moreover, these theories provide a way to determine $\tilde g(r)$ for any
simple system in the same way as in theory of simple liquids where the HS
system is used as reference system\cite{HM}. An alternative and fresh method
is to apply directly the present approach. It can be proposed a parametric
form of $\tilde g(r)$ which includes all the relevant physics. 
Extending the approach to systems with significant
compressibility, the two parameters of each peak, $\alpha$ and $R_i$, should
differ from their homologous of $\tilde g_0(r)$ in an amount which should
decrease quadratically with the distance. The normalization and equilibrium
equations should be enough to determine this decay and to reasonably describe
$\tilde g(r)$. Notice that the energy equation is another sum rule which can
be applied to these systems. The presence of defects, vacancies and
intersticials, would change the normalization of the peaks, in both $\tilde
g(r)$ and $\tilde g_0(r)$ which also would depend on the distance. For this
contribution, however, one should expect an exponential decay as the defects
would behave as a kind of fluid inside the solid which induce a short-range
correlation into $\tilde g(r)$.

\acknowledgments
We thank P. Tarazona for helpful discussions. This work has been supported by
the Direcci\'{o}n General de Investigaci\'{o}n Cient\'{\i}fica y T\'{e}cnica
of Spain under grant number $PB94-0005-C02$.

\begin{figure}
\caption{$\tilde g(r)$ at $\eta=0.52$. Solid line is the prediction of the
present theory using theoretical data from GELA functional approach. Dotted
line corresponds to the theoretical predictions using data from a
hypothetical exact theory. Triangles are Monte Carlo results from Choi {\it
et al.}. The inset shows details of the first peak ${\tilde g}_{1}(r)$.}
\end{figure}

\begin{figure}
\caption{As Figure 1. but for $\eta=0.71$.}
\end{figure}

\begin{figure}
\caption{Parameter $r_1$ (lower curve) and $<r>$ for the first neighbour
(upper curve) as a function of the mean density predicted by the theory.
Triangles from simulation data.}
\end{figure}

\end{document}